\documentclass[12pt, preprint]{aastex}

\newcommand {\hii}{H\,{\sc ii}}
\newcommand {\hi}{H\,{\sc i}}
\newcommand {\ha}{H$\alpha$}
\newcommand {\rah}{$^{\rm h}$}
\newcommand {\ram}{$^{\rm m}$}

\newcommand {\rosat}{{\it ROSAT\/}}

\newcommand {\cf}{{\em cf.}}
\newcommand {\him}{10$^6$~K}
\newcommand {\wim}{10$^4$~K}

\newcommand {\oqkev}{$1\over4$~keV}

\newcommand {\NH}{\mbox{$N_{\rm H}$}}        
\newcommand {\nhi}{\mbox{$N_{\rm H\,I}$}}    

\newcommand {\csam}{counts~s$^{-1}$~arcmin$^{-2}$}
\shorttitle{Diffuse X-ray Emission from the LMC}
\shortauthors{Points et al.}

\begin{document}


\title{Large Scale Diffuse X-ray Emission from the Large Magellanic Cloud}

\author{S. D. Points, Y.-H. Chu}
\affil{Astronomy Department, University of Illinois at Urbana-Champaign,
       Urbana, IL 61801}
\email{points@astro.uiuc.edu, chu@astro.uiuc.edu}


\author{S. L. Snowden\altaffilmark{1}}
\affil{NASA/Goddard Space Flight Center, Greenbelt, MD 20771}
\email{snowden@lheavx.gsfc.nasa.gov}

\and

\author{R. C. Smith}
\affil{Cerro Tololo Inter-American Observatory, Casilla 603, La Serena, Chile}
\email{csmith@noao.edu}

\altaffiltext{1}{Universities Space Research Association}

\begin{abstract}

X-ray mosaics of the Large Magellanic Cloud (LMC) taken with the
\rosat\ Position Sensitive Proportional Counter (PSPC) have revealed
extensive diffuse X-ray emission, indicative of hot $\ge 10^6$~K gas 
associated with this irregular galaxy on scales from $\sim 10$~pc 
to $\ge 1000$~pc.  We have selected regions of large-scale ($d \ge 
600$~pc) diffuse X-ray emission, such as supergiant shells, the LMC
Spur, and the LMC Bar, and examined the physical conditions of the 
hot gas associated with them.  We find that for these objects the 
plasma temperatures range from $kT \sim 0.15 - 0.60$~keV and the 
derived electron densities range from $n_e \sim 0.005 - 0.03$~cm$^{-3}$. 
Furthermore, we have examined the fraction of diffuse X-ray emission 
from the LMC and compared it to the total X-ray emission.  We find 
that discrete sources such as X-ray binaries and supernova remnants (SNRs)
account for $\sim 41\%$ and $\sim 21\%$ of the X-ray emission from the 
LMC, respectively.  In contrast, diffuse X-ray emission from the field 
and from supergiant shells account for $\sim 30\%$ and $\sim 6\%$ of
the total X-ray emission, respectively.

\end{abstract}



\section{Introduction}

X-ray images of the Large Magellanic Cloud (LMC) produced by
mosaicking pointed observations made with the {\em ROSAT\/} Position
Sensitive Proportional Counter (PSPC; Snowden \& Petre 1994) have
revealed extensive X-ray emission associated with this irregular
galaxy.  The PSPC mosaic of the LMC clearly exhibits diffuse X-ray
emission on scales of $\sim10$ -- 3000~pc, in addition to discrete
point sources such as foreground stars, LMC X-ray binaries, and
background active galaxies.  The diffuse X-ray emission is indicative
of the presence of hot ($\ge 10^6$~K) interstellar gas.  This hot gas
component, corresponding to the hot ionized medium (HIM) component of
McKee \& Ostriker's (1977) multi-phase interstellar medium (ISM), is
most likely shock-heated by supernova blasts and the fast stellar
winds of massive stars.  Some of this diffuse X-ray emission is
associated with interstellar structures such as \hii\ regions,
supernova remnants (SNRs), superbubbles, and supergiant shells, but
some of the diffuse emission appears to be unbounded by any
interstellar structure.

The LMC offers an ideal site to study the nature and origin of the
HIM, as well as its relationship to the other phases of the ISM.  At a
distance of 50~kpc, the LMC is near enough so that point sources can
be unambiguously distinguished from small diffuse sources, e.g., SNRs,
thus a wide range of diffuse X-ray sources can be analyzed.  With an
inclination angle of 30$^\circ-40^\circ$ and a foreground extinction
of A$_{\rm V} \sim 0.2 - 0.8$ mag (Westerlund 1997), the LMC can be
viewed with minimal confusion along the line-of-sight so that the
relationship among all phases of the ISM and the underlying stellar
population can be studied.

As such, much work has been performed to understand the nature and
origin of the diffuse X-ray emission from the LMC on all scales.
Williams et al. (1999) have recently completed an atlas of 31 SNRs in
the LMC to examine the physical conditions of the \him\ gas on the
micro-scale ($\sim 10$~pc).  Dunne, Points, \& Chu (2001) have
examined the physical conditions of the hot gas interior to a sample
of 13 X-ray bright superbubbles on the meso-scale ($\sim 100$~pc).
This paper addresses the physical properties of the $\ge 10^6$~K on
the macro-scale ($\sim 1000$~pc).

This paper is organized as follows: Section 2 describes the
observations used in this study and their reduction.  The distribution
of the hot ($\ge 10^6$~K) plasma and its relationship to the warm
($\sim 10^4$~K) gas on a global scale and for individual objects is
discussed in \S 3.  In \S 4 we determine the physical conditions of 
the $\ge 10^6$~K gas.  We summarize our results in \S 5.

\section{Observations and Data Reduction}

The X-ray mosaics of the LMC were obtained with the \rosat\ X-ray
telescope, using the PSPC which has a 2\arcdeg\ diameter
field-of-view.  The data were retrieved through the \rosat\ public
archive\footnote{http://heasarc.gsfc.nasa.gov/W3Browse/} at the High
Energy Astrophysics Science Archive Research Center (HEASARC) at
NASA's Goddard Space Flight Center.  Table~1 lists the observation
sequences that were used for the PSPC mosaics.  The table includes the
number of data sets for each sequence (the observation of individual
targets was often broken into multiple segments, some of which
required separate processing), pointing direction, exposure time, and
target name.  The data reduction was accomplished using the Extended
Source Analysis Software (ESAS)
package\footnote{ftp://legacy.gsfc.nasa.gov/rosat/software/fortran/sxrb/}
(Snowden \& Kuntz 1998; Kuntz \& Snowden 1998), which is also
available through the HEASARC.  The final PSPC mosaics are cast in the
zenith equal-area azimuthal (ZEA; Greisen \& Calabretta 1996)
projection, which is similar over this solid angle to tangential
projection.  This projection has a field center of $\alpha_{2000} =$
5\rah 28\ram, $\delta_{2000} = -67\arcdeg 15'$, a pixel size of
40$''$, and covers 51.27 square degrees.  We present the PSPC mosaic
of the LMC in the R4 to R7 (0.44 -- 2.04~keV; see Table~2) energy band
and the exposure map for the R5 energy band (0.56 -- 1.21~keV) in 
Figures~1 and 2, respectively.

Reduction of the PSPC data followed the procedures outlined by Snowden
et al. (1994) and demonstrated by Snowden \& Petre (1994).  In fact,
the PSPC mosaic presented in Figure~1 includes a reprocessing of data
presented by Snowden \& Petre (1994). This new mosaic was made from
129 pointed PSPC observations, whereas the previous mosaic was made
from $\sim 40$ pointed PSPC observations.  The individual observations
were screened for anomalous background conditions, had their residual
non-cosmic background components modeled and subtracted, and had their
vignetted and deadtime-corrected exposures calculated.  Only after all
of the individual observations were reduced in six
statistically-independent energy bands (see Table~2) were they cast
into the mosaics.  Because the analysis software cannot correct for
zero-level offsets (a constant component of the long-term enhancement
in the non-cosmic background, see Snowden et al. 1994), the mosaicking
process must correct for the relative offsets between overlapping
fields.  A single-value deconvolution algorithm was applied to a
system of equations comprising the count rates for all overlaps
between separate observations to determine a best fit for all offsets
simultaneously.  A final constraint was added so that the sum of the
offsets were equal to zero.  Individual observations were then
adjusted by multiplying the fitted offset by the exposure map to
produce an offset count image, which was then subtracted from the
count image.  This procedure was run separately for each energy band.
Table~3 lists the average exposure time, the total counts, the total
background counts, and the average intensity of each band.  The true
source counts can be seen to completely dominate the non-cosmic
background and the average exposures are larger by more than an order
of magnitude than that of the \rosat\ All Sky Survey coverage.

\section{Distribution of the Large Scale Diffuse X-ray Emission}

In this section we first describe the global distribution of the
diffuse X-ray emission toward the LMC and compare it with the spatial
distribution of $10^4$~K ionized gas in the LMC.  Based upon the X-ray
images, we show that the X-ray emission most likely has a hot $\ge
10^6$~K origin.  From these images, we select regions of large scale
diffuse X-ray emission that have, in most cases, optical counterparts.
Finally, we examine the relationship between the hot and the warm
ionized gas for the individual structures.

\subsection{Global Distribution of the X-ray Emitting Plasma}

As seen in Figure~1, the LMC clearly exhibits extensive diffuse
emission in the R4 to R7 energy band.  The diffuse X-ray emission in
the LMC occurs on physical scales from $\sim 10$~pc to $> 1000$~pc.
Observations of the LMC with the \rosat\ High Resolution Imager (HRI)
do not show evidence that the large scale diffuse X-ray emission is
produced by a population of point sources that were unresolved by the
PSPC, indicating a hot ($\ge 10^6$~K) gas origin (Chu \& Snowden
1998).  The X-ray surface brightness of the large-scale diffuse X-ray
emission is non-uniform across the LMC.  It varies from \hbox{$\sim
3.5 \times 10^{-4}$~\csam} in the northwest to \hbox{$\sim 4.0 \times
10^{-3}$~\csam} in the southeast in the R4 to R7 band.  This range of
X-ray surface brightness is comparable to that measured for LMC
superbubbles (\cf\ Dunne et al.  2001), but is a factor of $\sim 100$
lower than that of LMC SNRs (\cf\ Williams et al. 1999).  The large 
scale diffuse X-ray emission most likely originates in the LMC because 
it has a higher surface brightness than the nominal off-LMC value of 
$\sim 2.5 \times 10^{-4}$~\csam\ in the R4 to R7 band mosaic (Snowden 
\& Petre 1994; Snowden 1999).

We have compared the PSPC X-ray mosaic of the LMC with the \ha\ Sky
Survey image (Gaustad et al. 1999) of the LMC (Figure~3) to identify
the $10^4$~K ionized gas counterparts, if any, of the large scale
diffuse X-ray emission.  As can be seen by comparing Figures~1 and 3,
some of the large scale diffuse X-ray emission is associated with LMC
supergiant shells (e.g., LMC\,2), further suggesting a hot gas origin.
Some of the diffuse X-ray emission, however, appears to be unbounded
by any interstellar structure (e.g., the Spur and the LMC Bar).  The
regions of large scale diffuse X-ray emission toward the LMC that we
have selected for study, as well as their positions, physical sizes,
and the measured Galactic and LMC \hi\ column densities, $N_{\rm
H\,I}^{\rm Gal}$ and $N_{\rm H\,I}^{\rm LMC}$, respectively, toward
them are presented in Table~4.  The designations of the regions of
large scale diffuse X-ray emission presented in Table~4 are taken to
correspond with names given to their optical counterparts, where
appropriate.

\subsection{Individual Objects}

Below we describe the spatial distribution of the $10^4$~K gas and
compare it with the distribution of the $\ge 10^6$~K gas for the
regions of large scale diffuse X-ray emission listed in Table~4.  We
first discuss the distribution of the $10^4$~K ionized gas and then
the $\ge 10^6$~K gas of the selected regions because it is often
useful to use the optical features as a reference for the X-ray
emission.

The \ha\ images of the regions have been extracted from the PDS scans
of the Curtis Schmidt plates of Kennicutt \& Hodge (1986).  The two
exceptions for this are (1) LMC\,1, which is not covered by the PDS
scans, and (2) the LMC Bar region, which covers more than one PDS scan
field-of-view.  For LMC\,1, the \ha\ image has been mosaicked from
images obtained by the Magellanic Cloud Emission-line Survey (MCELS;
Smith et al. 1999).  For the LMC Bar, the image has been extracted
from the \ha\ Sky Survey (Gaustad et al. 1999) image of the LMC.  The
X-ray images for the individual objects have been extracted from the
PSPC mosaic of the LMC (Figure~1).

We present the \ha\ and PSPC X-ray images of the individual objects in
Figures~4 -- 12.  The gray-scale levels among the images are not the
same so that the objects with lower X-ray and \ha\ surface brightness
can be seen more clearly.  To aid in the comparison between the
$10^4$~K and $10^6$~K gas for each object we have plotted X-ray
contours at the 3, 5, 10, 15, and 20-$\sigma$ level above the local
background.  Because the selected regions cover a large spatial
extent, smaller scale X-ray sources (e.g., X-ray binaries, SNRs, and
superbubbles) also lie projected along the line-of-sight.  Therefore,
we have also plotted additional contour levels at 50, 100, 150, and
200-$\sigma$ level above the mean local background.  These contours
are plotted as dashed lines.  Small-scale features mentioned in the
text such as \hii\ regions, SNRs, and superbubbles have also been 
labeled.

\subsubsection{LMC\,1}

The \ha\ image of LMC\,1 (Figure~4\,a) shows a very coherent shell
structure with a diameter of $\sim 700$~pc.  The northern and eastern
filaments of LMC\,1 are better defined than the filaments in the west
and south.  A single \hii\ region (N\,13) lies projected toward the
southeastern side of LMC\,1.  Diffuse X-ray emission is detected
interior to the \ha\ filaments defining LMC\,1.

The diffuse X-ray emission from within LMC\,1 seems to fill the shell
in the area covered by the PSPC mosaic (Figure~4\,b).  The diffuse
X-ray emission from within LMC\,1 peaks in the western portion of the
shell and along the southern rim.  Some of the diffuse X-ray emission
extends beyond the ionized filaments in the southwest, suggesting a
blowout of the hot gas.  High resolution \hi\ aperture synthesis
images of LMC\,1 (Kim et al. 1999) show \hi\ filaments extending beyond 
the X-ray emission region, indicating that the hot gas is confined by the
\hi\ gas.

\subsubsection{LMC\,2}

The distribution of $10^4$~K ionized gas toward LMC\,2 is revealed in
the \ha\ image (see Figure~5\,a).  This image shows long filaments in
the north and the east and shorter filaments in the south.  Together,
these filaments suggest a coherent shell structure with a diameter of
$\sim 900$~pc.  LMC\,2 is bordered on its western edge by the most
active star formation complex in the LMC: N\,157 (30~Doradus), N\,158,
N\,160, and N\,159.  In addition to this ridge of star formation, many
other \hii\ regions lie along the periphery or are projected interior
to LMC\,2.  Two SNRs are projected near the periphery of LMC\,2:
DEM~L~299 at the northern border and DEM~L~316 in the southeast
(Mathewson et al.  1983; Williams et al. 1997).  Diffuse X-ray
emission is detected interior to the ionized filaments that define the
boundary of LMC\,2.

The spatial distribution of X-ray emission and the physical properties
of the hot gas toward LMC\,2 have been extensively studied (Wang \&
Helfand 1991; Points et al. 1999; Points et al. 2000); therefore, we
summarize those results here.  LMC\,2 has the highest X-ray surface
brightness of all the supergiant shells in the LMC; the peak of the
diffuse emission is approximately 35 times higher than the off-LMC
background.  The diffuse X-ray emission from within LMC\,2 is confined
by the filaments to the north and east, but extends beyond the
southern filaments in a bright X-ray Spur (see Figure~11 and \S 3.2.8).
The diffuse X-ray emission toward LMC\,2 is non-uniform.  A bright
X-ray arc is seen in the southwest that appears to extend from N\,158
to N\,159, centered on N\,160.  A region of lower X-ray surface
brightness lies between the X-ray arc and a region of bright emission
in the northeast.  The bright X-ray arc in the southwest has been
suggested as a blowout of hot gas from N\,160 into LMC\,2.  The region
of low X-ray surface brightness ($\alpha_{2000} = $ 5\rah 45\ram,
$\delta_{2000} = -69^{\circ} 25'$) is likely caused by absorption from
an \hi\ cloud on the front side of LMC\,2.  The bright northern region
is probably energized by an SNR shock interior to LMC\,2 hitting the
shell wall.  As seen in Figure~5\,b, discrete X-ray emission sources
are located toward LMC\,2.  These sources include 30~Doradus, SNRs 
(i.e., SNR~0540$-$69.3, DEM~L~299, \& DEM~L~316) and a high mass X-ray 
binary (i.e., LMC~X-1).

\subsubsection{LMC\,3}

The \ha\ image of LMC\,3 (Figure~6\,a) shows an intricate filamentary
shell structure to the northwest of the 30~Doradus star formation
complex with a diameter $\sim 1000$~pc.  The ionized filaments do not
have the coherence of those seen in LMC\,1 and LMC\,2, but are
suggestive of a complete shell structure.  Several \hii\ regions lie
along the periphery of LMC\,3 and appear to be connected by the
ionized filaments.  Diffuse X-ray emission is detected toward LMC\,3.

The diffuse X-ray emission toward LMC\,3 covers the interior of the
supergiant shell (see Figure~6\,b).  The X-ray surface brightness of
the diffuse emission is non-uniform. The diffuse X-ray emission peaks
along the eastern rim of LMC\,3.  Several local depressions in the
X-ray surface brightness are also seen in the PSPC mosaic toward
LMC\,3.  Two of these minima are coincident with \hii\ regions
projected toward LMC\,3: N\,148 (DEM~L~227) along the northern rim and
DEM~L~210 toward the center.  These two depressions are more
pronounced in the R4$+$R5 energy band than in the R6$+$R7 energy band,
indicative of absorption.  Therefore, it is likely that N\,148 and 
DEM~L~210 lie on the front side of LMC\,3.  In addition to the
large scale diffuse X-ray emission associated with LMC\,3, discrete
X-ray sources such as foreground stars, SNRs (e.g., SNR 0528$-$692), 
and superbubbles (e.g., N\,144) are seen in Figure~6\,b.

\subsubsection{LMC\,4} 

LMC\,4 is the largest optically identified supergiant shell in the LMC
with a dimension of 1400~pc $\times$ 1000~pc.  LMC\,4 consists of fragmented
ionized gas filaments that seem to connect the numerous \hii\
regions along its periphery, suggesting a coherent shell structure
(Figure~7\,a).  Diffuse X-ray emission is detected within LMC\,4.

The physical conditions of the hot ($\ge 10^6$~K) gas interior to
LMC\,4 have been reported by Bomans, Dennerl, \& K\"urster (1994).
The PSPC mosaic of LMC\,4 (Figure~7\,b) shows very low surface
brightness emission.  LMC\,4 has a a low foreground \hi\ column
density of \nhi\ $= 1.26 - 1.84 \times 10^{21}$~cm$^{-2}$.  Therefore, the
absence of strong X-ray emission detected toward LMC\,4 is not caused
by absorption.  In general, the X-ray emission from the interior of
LMC\,4 appears limb-brightened, but the brightest diffuse X-ray
emission in the west and south rims is only a factor of $\sim 2$
brighter than the X-ray emission from the center of the supergiant
shell.  The X-ray emission from the western side of LMC\,4 is not
associated with any \ha\ emission features and is probably gas
interior to LMC\,4.  The diffuse X-ray emission detected toward the
southern boundary of LMC\,4, however, is more confusing.  This
emission is most likely associated with the superbubbles that border
LMC\,4 along the southern rim, such as N\,51 and N\,57.  Discrete
X-ray sources, such as an X-ray binary (i.e., LMC~X-4), and SNRs
(i.e., N\,49, N\,49\,B, DEM~L~241, \& N\,63\,A) are also seen in the
PSPC mosaic image.

\subsubsection{LMC\,5}

The \ha\ image of LMC\,5 (Figure~8\,a) shows faint, irregular ionized
filaments with a diameter of $\sim 800$~pc to the northwest of LMC\,4.
A long ionized filament ($\sim 300$~pc long) appears to bisect the
LMC\,5, giving the impression that LMC\,5 is comprised of two distinct
shells.

The diffuse X-ray emission from LMC\,5 covers most of its interior
(see Figure~8\,b).  The brightest diffuse X-ray emission from LMC\,5
is located to the south of the \ha\ filament that divides the shell.
The bright diffuse X-ray emission at the south rim cannot be
unambiguously associated with LMC\,5, however, because it lies
projected toward an \hii\ region (N\,48).  A region of lower X-ray
surface brightness is seen along the western rim of the shell,
coincident with the \ha\ filaments.  The lowered X-ray surface
brightness may be the result of the ionized gas absorbing the X-ray
emission from inside the supergiant shell.  The strong X-ray sources
to the east of LMC\,5 are the SNRs, N\,49 and N\,49\,B.  The X-ray
sources to the west of the LMC\,5 are two foreground Galactic late-type stars.

\subsubsection{LMC\,6}

The warm ionized shell of LMC\,6 is very faint and circular in
appearance, with a diameter of $\sim 600$~pc (see Figure~9\,a).  It is
the smallest supergiant shell in the LMC (Meaburn 1980).  The surface
brightness of the ionized filaments is highest toward the \hii\
regions N\,91 and N\,92 along the northern and southern boundaries,
respectively.  In addition to these \hii\ regions, the SNR N\,86 is
projected on the western rim of LMC\,6.

The spatial distribution of the X-ray emitting plasma from LMC\,6
appears limb-brightened (see Figure~9\,b).  LMC\,6 has a very low
X-ray surface brightness, $\la 2$ times the off-LMC background.  The
peak diffuse X-ray emission toward LMC\,6 lies toward N\,91 and may be
associated with the \hii\ region.  The bright, discrete X-ray sources
toward LMC\,6 are the N\,86 SNR and a late-type Galactic star.

\subsubsection{LMC\,10 (DEM~L~268)}

The large scale structure designated as LMC\,10 was tentatively
identified as a supergiant shell candidate by Meaburn (1980).  These
faint ionized filaments lie to the north of 30~Doradus and LMC\,3
(Figure~10\,a).  The \ha\ morphology of the LMC\,10 show the filaments
to be pointing away from 30~Doradus and LMC\,3, particularly in the
area toward N\,148 along the northeastern rim of LMC\,3.

The PSPC mosaic of LMC\,10 (Figure~10\,b) shows an arc of X-ray
emission that stretches from the northern edge of LMC\,3 and may
continue toward the southern edge of LMC\,4.  The X-ray emission from
this arc has a fairly high surface brightness, with a peak
approximately 5 times the off-LMC background.  The extent of the X-ray
emission of LMC\,10 cannot be measured in our PSPC images because of
incomplete coverage of the mosaic toward its central region.

\subsubsection{LMC Spur}

The X-ray Spur in the LMC lies to the south of the supergiant shell
LMC\,2 (see Figure~1).  The \ha\ image of the Spur (Figure~11\,a)
shows several \hii\ along its western edge, but does not reveal any
\wim\ ionized filaments encompassing it.  Thus, its boundaries and
physical size are defined solely from the PSPC mosaic.  Even though
little $10^4$~K ionized gas is detected toward the Spur, other phases
of the ISM are present.  Observations of the $^{12}$CO $J = 1
\rightarrow 0$ emission line at 2.6~mm made with the NANTEN 4-m radio
telescope (Fukui et al. 1999) reveal that the Spur is bordered on the
west by a ridge of molecular gas.  The \hii\ regions along the western
edge are superposed on this molecular ridge.

The PSPC mosaic image (Figure~11\,b) reveals that the Spur has a
physical size of $\sim 900$~pc, comparable to LMC\,2.  The Spur has
the second highest average X-ray surface brightness of all the objects
in this study.  It has been suggested that the Spur represents a
blowout of the hot gas interior to LMC\,2 into the halo of the LMC
(Wang \& Helfand 1991) although high-resolution optical spectra toward
the Spur do not reveal any high-velocity gas that would be indicative
of an outflow (Points et al. 1999).  Very little structure or
variation in the X-ray surface brightness is observed toward the Spur.
This is remarkable because LMC\,2 shows X-ray surface brightness
variations on smaller scales that are caused by differential
absorption (Points et al. 1999; Points et al. 2000).  The smooth
distribution of X-ray emission from the Spur may imply a very uniform
foreground absorption.  The western boundary of the Spur is coincident
with the ridge of molecular gas.  This molecular material could be
absorbing X-ray emission from the Spur and be responsible for the
sharp western edge of the X-ray emission.

\subsubsection{LMC Bar}

The LMC Bar primarily consists of an intermediate-age stellar
population upon which many \hii\ regions are superposed (see
Figure~12\,a) such as N\,113, N\,114, N\,119, and N\,120.  The X-ray
bar of the LMC is the largest region of diffuse X-ray emission
reported here with a major axis $\sim 3000$~pc and a minor axis $\sim
1000$~pc.

The PSPC mosaic (Figure~12\,b) shows bright large-scale diffuse gas
from the bar region of the LMC.  The HRI survey of the LMC (Chu \&
Snowden 1998) does not show large population of point sources that 
are unresolved by the PSPC.  We note, however, that the \rosat\ HRI 
survey of the LMC has a detection limit of $\sim 1 \times 10^{34}$~erg~s$^{-1}$ 
for point sources (Chu \& Snowden 1998).  Thus, some of the X-ray 
emission from the LMC Bar could consist of point sources that were 
unresolved by the PSPC, but not detected by the HRI because of its 
high background.  If the X-ray emission from the LMC Bar is produced
by coronal emission from unresolved late-type stars in the LMC Bar, 
the stellar density of this population would be $\sim 120$~pc$^{-3}$,
which is unrealistically high.  Therefore, the X-ray emission from 
the LMC Bar appears to be truly diffuse.  In general, the X-ray 
emission is not spatially correlated with the star forming regions 
superposed on the LMC Bar.  Discrete X-ray sources such as SNRs (e.g., 
SNR~0519$-$690, SNR~0520$-$694, N,120, and N\,132~D) and foreground 
stars are seen projected toward the LMC Bar.

\section{Discussion}

\subsection{Physical Properties of the Hot Gas}

Before the physical conditions of the $\ge 10^6$~K gas can be
determined, we must first excise the discrete X-ray emission sources,
such as foreground Galactic stars, LMC X-ray binaries, SNRs, and
superbubbles, and background active galactic nuclei, from all of the
PSPC mosaic band images.  For this purpose, we used the catalog of
\rosat\ PSPC X-ray sources toward the LMC that was compiled by Haberl
\& Pietsch (1999) to obtain the positions of the discrete sources.
After the contaminating sources were removed from the mosaics, the
background-subtracted PSPC count rates are determined for the regions
of large scale diffuse X-ray emission in each PSPC band.  The count
rates in the individual bands were then summed to obtain count rates
in the R4 $+$ R5 (0.44 -- 1.21~keV) and R6 $+$ R7 (0.73 -- 2.04~keV)
energy bands.  The angular size of the extraction regions, net source
count rates in the PSPC energy bands and the ${\rm (R6 + R7)} \over
{\rm (R4 + R5)}$ ratio for the regions of large scale diffuse X-ray
emission are presented in Table~5.

The measured X-ray surface brightness of the hot gas is dependent upon
the temperature of the thermal plasma and the foreground absorption.
If the foreground absorption can be independently measured, the
intensity ratio between the different bands can be used to determine
the temperature of the emitting gas.  We use a Raymond \& Smith (1977)
thermal plasma emission model with 40\% solar abundance\footnote{We
discuss our selection of 40\% solar abundance plasma emission models
below.} to interpret the ${\rm (R6 + R7)} \over {\rm (R4 + R5)}$ band
intensity ratios of the selected regions as an effective temperature
of the hot plasma associated with them.  We use measurements of the
Galactic and LMC \hi\ column densities from the data presented by
Dickey \& Lockman (1990) and Rohlfs et al. (1984) respectively, to
determine lower and upper limits on the total foreground absorption
column to the regions of interest in the LMC.  As seen in Table~4, the
Galactic \hi\ column density, $N_{\rm H\,I}^{\rm Gal}$, ranges from
$\sim 4 - 7 \times 10^{20}$~cm$^{-2}$.  Arabadjis \& Bregman (1999)
have shown that for column densities \NH\ $> 5 \times
10^{20}$~cm$^{-2}$, the total X-ray absorption column is nearly double
the \hi\ column density because of the contribution of molecular gas.
Although molecular gas does contribute to the total absorption column
above column densities $\sim 5 \times 10^{20}$~cm$^{-2}$, it is
neither a factor of 2 uniformly over the sky, nor a step function at
$5 \times 10^{20}$~cm$^{-2}$.  Thus, the contribution to the total
absorption column by the Galactic ISM probably ranges from 1 to 2
times $N_{\rm H\,I}^{\rm Gal}$.  Continuing this approximation to the
LMC absorption column is more difficult.  Measurements of the LMC \hi\
column density sample material that is both in front of and behind the
regions of interest.  We use the simplifying assumption that half of
the neutral atomic gas is foreground to the regions and that half is
background.  Thus, the LMC component of the total absorption column
density is ${1 \over 2} \times 2 N_{\rm H\,I}^{\rm LMC} = N_{\rm
H\,I}^{\rm LMC}$.  The total absorption column is between $N_{\rm
H\,I}^{\rm Gal} + N_{\rm H\,I}^{\rm LMC}$ and $2 N_{\rm H\,I}^{\rm
Gal} + N_{\rm H\,I}^{\rm LMC}$.  We use both of these limiting values
of the foreground absorption to determine the physical conditions of
the $\ge 10^6$~K gas in the regions of interest.
Figure~13 shows a plot of the PSPC ${\rm (R6 + R7)} \over {\rm (R4 + R5)}$ band ratio 
versus \NH\ for a set of 40\% solar abundance Raymond \& Smith (1977) plasma 
emission models with temperatures ranging from ${\rm Log ({T \over K})} = 
6.2 - 6.9$ (or $kT = 0.14 - 0.68$~keV).  For the selected regions of large scale 
diffuse X-ray emission in the LMC, the plasma temperature varies from $kT 
\sim 0.16 - 0.60$~keV.
As seen in Figure~13, the errors in the ${\rm (R6 + R7)} \over {\rm (R4
+ R5)}$ band ratios imply that the supergiant shell LMC\,2 could have
a plasma temperature ranging from $kT = 0.22$ to 0.68~keV if the
foreground absorption is $4.4 \times 10^{21}$~cm$^{-2}$ for a
Raymond \& Smith (1977) 40\% solar abundance plasma emission model.
This range in plasma temperature for the hot gas interior to LMC\,2
is similar to the 90\% confidence limits on the plasma temperature
that were determined by thermal plasma model fits to \rosat\ PSPC
spectra of LMC\,2 (Points et al. 2000). 

From the determination of the plasma temperatures and the measured
foreground absorption, we use
PIMMS\footnote{http://heasarc.gsfc.nasa.gov/docs/software/tools/pimms.html}
(Mukai 1993) to calculate the unabsorbed X-ray fluxes, and hence X-ray
luminosities, of the regions in the R4 to R7 energy band.  Given a hot
gas filling factor, $f$, volume, $V$, X-ray luminosity, $L_X$, and
emissivity, $\Lambda$, the electron density of the hot plasma will be
$$n_e = (1.1 L_X)^{1/2} (\Lambda V f)^{1/2},$$
if $n_e = 1.1 n_{\rm H}$.  For the selected regions, we assume that
the volume of the gas is equal to the product of the surface area and
the path length through the gas.  We adopt a path length comparable to
the width of the region in question.  We derive the electron densities
for the large scale diffuse X-ray emission regions using the
calculated X-ray luminosities and emissivities and present them in
Table~6.  The average electron densities in these regions varies from
$\sim 0.004$~cm$^{-3}$ in LMC\,4 to $\sim 0.03$~cm$^{-3}$ in the LMC 
Spur.

The preceding derivation of the physical conditions of the hot gas
associated with the large scale diffuse X-ray emission in the LMC is
not without its caveats.  The major uncertainties in our calculations 
do not arise from uncertainties in the measured count rates, but are
attributed to the assumptions we have made concerning the metallicity 
of the X-ray emitting gas, the geometry of the emitting region and 
the foreground absorption.  For example, the canonical chemical
abundance of the ISM in the LMC is 30\% of the solar value (Russell \&
Dopita 1992).  In this work, however, we have determined the physical
conditions of the hot gas using a Raymond \& Smith (1977) thermal
plasma with 40\% solar abundance because PIMMS does not have a grid of
30\% solar abundance Raymond \& Smith (1977) models.  Therefore, we
have calculated the plasma temperatures, X-ray luminosities, and
electron densities of the selected regions using a Raymond \& Smith
(1977) thermal plasma with 20\% solar abundance to investigate the
importance of the chemical abundance on the physical conditions of the
hot gas.  This work shows that the mean percent difference in plasma
temperatures between Raymond \& Smith (1977) models with 20\% and 40\%
solar abundances is $\sim 4\%$ for the individual objects, but that
the mean percent differences in the X-ray fluxes and electron
densities are $\sim 33\%$ and $\sim 20\%$, respectively.
Another complicating factor in calculating the physical properties of
the 10$^6$~K gas is the uncertainty in the three dimensional geometry, 
or volume, of the individual regions.  Therefore, we simply take the volume 
to be the product of the projected surface area and the path length through 
the gas where the path length is equal to the width of the structure.
Because the depth through these structures is uncertain, an increase
(decrease) of the path length, $l$, by a factor of two, would decrease
(increase) our determination of the electron density by a factor of
$\sqrt 2$.
The final assumption that we have made is that the foreground
absorption column toward the individual regions of interest is uniform
and can be represented as a combination of the Galactic and LMC \hi\
column densities.  We have shown in \S 3.2 that some of the
fluctuations in the X-ray surface brightness toward the individual
objects are associated with differential absorption.  Therefore, the
assumption of uniform absorption across each region is clearly not
correct if we want to investigate the variations in plasma temperature
and electron density across each region of interest.  Thus, we only
can determine the average physical conditions of the hot gas in the
regions.

\subsection{Fraction of Diffuse Emission to Total Emission}

In addition to determining the physical properties of the $\ge 10^6$~K
gas in our sample of selected objects, we have also investigated the
contribution of these objects to the total X-ray emission observed
from the LMC.  This comparison will allow us to better understand the
X-ray emission from more distant Magellanic Irregular galaxies with
star formation rates of 0.26 M$_{\odot}$~yr$^{-1}$ (Kennicutt et
al. 1995).  As previously discussed, the X-ray emission from the PSPC
mosaics is comprised of emission from: (1) AGN and background
galaxies, (2) foreground stars, (3) the soft X-ray background, and (4)
LMC sources (e.g., X-ray binaries, SNRs, superbubbles, supergiant
shells, and field emission).  Therefore, in order to determine the
amount of X-ray emission that is intrinsic to the LMC, we must
determine and subtract the contributions of the background galaxies,
foreground stars, and the soft X-ray background from the PSPC mosaic.
Below we discuss the removal of these different sources from PSPC
mosaics and determine the X-ray emission that is attributed to sources
in the LMC.

The first step in calculating the diffuse X-ray emission from the LMC
is determining the soft X-ray background toward the LMC in the R1 to
R7 (0.1 -- 2.04~keV) energy band and subtracting it from the PSPC
mosaics.  To determine the soft X-ray background in this energy range
we have created two broadband X-ray images of the LMC: the first in
the R4 to R7 (0.44 -- 2.04~keV) energy band (see Figure~1) and the
second in the R1 to R2 (0.1 -- 0.28~keV) energy band (see Figure~1 of
Snowden 1999).  The off-LMC background in the R4 to R7 band has been
previously measured by Snowden \& Petre (1994) to be $\sim 2.5 \times
10^{-4}$~\csam.  Unfortunately, there are no published values of the
off-LMC background in the R1 to R2 band; therefore, we need to measure
the value of the background directly from the R1 to R2 mosaic of the
LMC.  To accomplish this task, we have determined the mean count rate
in 10 regions, each having an angular extent of 10 arcmin$^{2}$, along
the periphery of the R4 to R7 PSPC mosaic.  The average background
measured in these 10 regions is $\sim 2.4 \times 10^{-4}$~\csam, in
good agreement with the value reported by Snowden \& Petre (1994).
Somewhat confident that these 10 regions are representative of the
off-LMC background in the R4 to R7 band, we use them to determine the
off-LMC background to be $\sim 4.2 \times 10^{-4}$~\csam\ in the R1 to
R2 band\footnote{The soft X-ray background in the R1 to R2 energy band
has been measured to be $\sim 1.04 \times 10^{-3}$~\csam\ toward the
north Galactic Pole (Kuntz \& Snowden 2000).  Snowden et al. (1998)
note that the soft X-ray emission in the northern hemisphere is $\sim
2$ times stronger than in the southern hemisphere in this energy
range.  The value we determined for the background in the R1 to R2
band is within 20\% of the expected background of $\sim 5 \times
10^{-4}$~\csam.}.  After the off-LMC backgrounds in the R4 to R7 and
R1 to R2 energy bands are subtracted, we combine the resulting images
to make a mosaic of the LMC in the R1 to R7 energy band.

Now that the background has been subtracted from the PSPC mosaic in
the R1 to R7 energy band, we use the list of X-ray sources toward the
LMC compiled by Haberl \& Pietsch (1999) to determine the contribution
of discrete, point-like sources (e.g., AGN, foreground stars, X-ray
binaries, supersoft X-ray sources, SNRs, and unclassified X-ray
sources) to the total emission in the PSPC mosaics.  We have also
taken the list of superbubbles in the LMC compiled by Dunne et
al. (2001) and the list of large scale structures discussed in \S 3 to
determine their contribution to the total emission in the PSPC mosaic.
After the contributions from all of these sources are subtracted from
the total X-ray emission in the background-subtracted PSPC mosaic, we
are left with the diffuse X-ray emission from the field of the LMC.
We present these results in Table~7.  As seen in Table~7, the largest
contribution to the X-ray emission from the LMC comes from X-ray
binaries and supersoft X-ray sources (41\%).  Surprisingly, the second
largest contribution to X-ray emission from the LMC is from diffuse
gas in the field (29\%).  This is slightly higher than the X-ray
emission produced by SNRs (21\% of the total) and three times the
amount of X-ray emission associated with supergiant shells, the Spur
and the Bar.  Diffuse X-ray emission from superbubbles makes a small
contribution to the total X-ray emission from the LMC (2\%).

As with any calculation, we must consider the factors that produce
uncertainties.  The largest source of error in the determination of of
total X-ray emission from the LMC, and hence the emission from diffuse
gas in the field is the value of the off-LMC X-ray background in the
R1 to R2 energy band.  If the off-LMC background in the R1 to R2 band
is one half of the X-ray background in the R1 to R2 band observed
toward the north Galactic Pole (see Kuntz \& Snowden 2000), then the
contribution of diffuse X-ray emission from the field to the total
X-ray emission from the LMC drops from 27\% to 20\% of the total.
This is primarily a consequence of the method by which we determined
the amount of diffuse X-ray emission from the field.  To determine the
count rate for the field emission we subtracted the emission from the
cataloged sources from the total emission in the background-subtracted
PSPC mosaic.  We attribute the difference between the total X-ray
emission in the LMC and the cataloged sources to be from diffuse gas
in the field.  Thus, by increasing the soft X-ray background in the R1
to R2 energy band, we decrease the total number of counts in the PSPC
mosaic, effectively reducing the amount of diffuse X-ray emission from
the field.  Another source of potential error in the determination of
the amount of X-ray emission intrinsic to the LMC is errors in the
count rates we determined from the PSPC mosaics for the cataloged 
superbubbles in the LMC (\cf, Dunne et al. 2001) and the objects listed 
in \S3.  To check the accuracy of our source extraction from the mosaics,
we have also determined the count rates of the cataloged SNRs and X-ray 
binaries in the LMC and compared these values with those determined by
Haberl \& Pietsch (1999).  The total count rates that we determined for
SNRs and X-ray binaries are within 5\% of those determined by Haberl \&
Pietsch (1999).  Therefore, we conclude that our method of extracting
source counts directly from the PSPC mosaics does not significantly affect
our determination of the total X-ray emission from the LMC and that 
the largest source of error lies in the determination of the off-LMC 
background.  Finally, there is also considerable uncertainty due to the 
foreground absorption to the LMC.  The Galactic \hi\ column densities 
toward the LMC are $\sim 6 \times 10^{20}$~cm$^{-2}$, corresponding to
4 -- 5 optical depths at \oqkev.  Thus, small errors in the assumed
foreground absorption can have large effects on the determined PSPC count 
rates.

\section{Summary}

The PSPC mosaics of the LMC provide an unprecedented and detailed
global view of the $\ge 10^6$~K gas in a galaxy.  The diffuse X-ray 
emission is non-uniform across LMC and shows a highly complex ISM 
with features on the scale from a few arcminutes to over several 
degrees.  Furthermore, these X-ray mosaics allow us to examine
the spatial distribution and physical conditions of diffuse X-ray
emission on scales larger than a single pointed PSPC observation
allows.

We have selected regions of large-scale X-ray emission in the LMC and
compared them with \ha\ images.  This comparison shows that while some
of the large scale diffuse X-ray emission is associated with
supergiant shells, some is also unbounded such as the LMC Spur, and
the LMC bar.  We have determined the physical properties of these
regions of large scale diffuse emission.  We find that the plasma
temperature varies from $\sim 0.2$ to $\sim 0.6$~keV and the X-ray
surface brightness varies from $\sim 0.07 $ -- $ \sim 5 \times 
10^{-14}$~~erg~cm$^{-2}$~s$^{-1}$~arcmin$^{-2}$.  The electron 
density of the hot gas in these regions varies from 0.003~cm$^{-3}$ to 
0.030~cm$^{-3}$.

We have determined the X-ray count rate from the LMC in the R1 to R7 
energy band.  Discrete sources, such as foreground stars, AGN, X-ray 
binaries, SNRs, and superbubbles, were removed.  From this we are
able to determine the X-ray count rate from the selected regions and
from the field.  We find that X-ray emission from the field of the
LMC contributes between 20\% to 27\% of the total X-ray emission from
the LMC, depending on how the X-ray background is removed.  The diffuse
X-ray emission from the field provides a significant contribution to 
the total X-ray luminosity of the LMC.  It is comparable to the amount
of X-ray emission associated with SNRs in the LMC and is a factor of 2 
to 3 times greater than the X-ray emission associated with supergiant 
shells, the Spur, and the Bar in the LMC.

\acknowledgments This research is supported by NASA grants NAG 5-8104 and
NAG 5-10011.  This research has made use of data obtained from the High 
Energy Astrophysics Science Archive Research Center (HEASARC), provided 
by NASA's Goddard Space Flight Center.  We gratefully acknowledge the
support of CTIO and all the assistance which has been provided in upgrading
the Curtis Schmidt telescope.  The MCELS is funded through the support
of the Dean B. McLaughlin fund at the Univ. of Michigan and through
NSF grant \#9540747.  The authors would also like to thank E. Schlegel
for his careful review of this manuscript.

\clearpage


\begin{figure}
\figurenum{1}
\caption{\rosat\ PSPC mosaic of the LMC in the R4 to R7 (0.44 -- 2.04~keV) 
energy band.  Darker regions indicate higher intensity.  The image has been 
logarithmic scaled with an intensity range that saturates at $5 \times 
10^{-3}$~\csam.  The regions discussed in \S 3.2 are marked.  The bar in the 
upper left corner indicates a length 500~pc, assuming a distance of 50~kpc 
to the LMC.}
\end{figure}


\begin{figure}
\figurenum{2}
\caption{Gray-scale map of the R5 (0.56  -- 1.21~keV) energy band 
exposure.  Darker regions indicate higher exposure.  The image has been
logarithmic scaled with an intensity range that saturates at 100~ksec.}
\end{figure}


\begin{figure}
\figurenum{3}
\caption{\ha\ Sky Survey image of the LMC (Gaustad et al. 1999).  The
image has been sky- and continuum-subtracted.  The white ``holes'' are
an artifact of the continuum subtraction.  The image has been logarithmic
scaled and covers the same region as Figure~1.  The individual regions 
discussed in \S 3.2 are marked.  The bar in the upper left corner indicates 
a length 500~pc, assuming a distance of 50~kpc to the LMC. }
\end{figure}


\begin{figure}
\figurenum{4}
\caption{{\bf (a)} MCELS \ha\ image of LMC\,1. {\bf (b)} \rosat\ PSPC
image of LMC\,1 in the R4 -- R7 (0.44 -- 2.04~keV) energy band.  {\bf 
(c)} and {\bf (d)} The same as {\bf (a)} and {\bf (b)} with X-ray contours
overlaid.  The solid contours in {\bf (c)} and {\bf (d)} are 3, 5, 10, 
15, and 20-$\sigma$ above the mean local background.  The dashed contours 
are 50, 100, 150, and 200-$\sigma$ above the mean local background.}
\end{figure}


\begin{figure}
\figurenum{5}
\caption{{\bf (a)} Curtis-Schmidt \ha\ PDS scan of LMC\,2 (Courtesy
R. Kennicutt). {\bf (b)} \rosat\ PSPC image of LMC\,2 in the R4 -- R7
(0.44 -- 2.04~keV) energy band. {\bf (c)} and {\bf (d)} The same as
{\bf (a)} and {\bf (b)} with X-ray contours overlaid.  The solid
contours in {\bf (c)} and {\bf (d)} are 3, 5, 10, 15, and 20-$\sigma$
above the mean local background.  The dashed contours are 50, 100,
150, and 200-$\sigma$ above the mean local background.}
\end{figure}


\begin{figure}
\figurenum{6}
\caption{{\bf (a)} Curtis-Schmidt \ha\ PDS scan of LMC\,3 (Courtesy
R. Kennicutt). {\bf (b)} \rosat\ PSPC image of LMC\,3in the R4 -- R7
(0.44 -- 2.04~keV) energy band. {\bf (c)} and {\bf (d)} The same as
{\bf (a)} and {\bf (b)} with X-ray contours overlaid.  The solid
contours in {\bf (c)} and {\bf (d)} are 3, 5, 10, 15, and 20-$\sigma$
above the mean local background.  The dashed contours are 50, 100,
150, and 200-$\sigma$ above the mean local background.}
\end{figure}


\begin{figure}
\figurenum{7}
\caption{{\bf (a)} Curtis-Schmidt \ha\ PDS scan of LMC\,4 (Courtesy
R. Kennicutt). {\bf (b)} \rosat\ PSPC image of LMC\,4 in the R4 -- R7
(0.44 -- 2.04~keV) energy band. {\bf (c)} and {\bf (d)} The same as
{\bf (a)} and {\bf (b)} with X-ray contours overlaid.  The solid
contours in {\bf (c)} and {\bf (d)} are 3, 5, 10, 15, and 20-$\sigma$
above the mean local background.  The dashed contours are 50, 100,
150, and 200-$\sigma$ above the mean local background.}
\end{figure}


\begin{figure}
\figurenum{8}
\caption{{\bf (a)} Curtis-Schmidt \ha\ PDS scan of LMC\,5 (Courtesy
R. Kennicutt). {\bf (b)} \rosat\ PSPC image of LMC\,5 in the R4 -- R7
(0.44 -- 2.04~keV) energy band. {\bf (c)} and {\bf (d)} The same as
{\bf (a)} and {\bf (b)} with X-ray contours overlaid.  The solid
contours in {\bf (c)} and {\bf (d)} are 3, 5, 10, 15, and 20-$\sigma$
above the mean local background.  The dashed contours are 50, 100,
150, and 200-$\sigma$ above the mean local background.}
\end{figure}


\begin{figure}
\figurenum{9}
\caption{{\bf (a)} Curtis-Schmidt \ha\ PDS scan of LMC\,6 (Courtesy
R. Kennicutt). {\bf (b)} \rosat\ PSPC image of LMC\,6 in the R4 -- R7
(0.44 -- 2.04~keV) energy band. {\bf (c)} and {\bf (d)} The same as
{\bf (a)} and {\bf (b)} with X-ray contours overlaid.  The solid
contours in {\bf (c)} and {\bf (d)} are 3, 5, 10, 15, and 20-$\sigma$
above the mean local background.  The dashed contours are 50, 100,
150, and 200-$\sigma$ above the mean local background.}
\end{figure}


\begin{figure}
\figurenum{10}
\caption{{\bf (a)} Curtis Schmidt \ha\ PDS scan of LMC\,10 (Courtesy
R. Kennicutt). {\bf (b)} \rosat\ PSPC image of LMC\,10 in the R4 -- R7
(0.44 -- 2.04~keV) energy band. {\bf (c)} and {\bf (d)} The same as
{\bf (a)} and {\bf (b)} with X-ray contours overlaid.  The solid
contours in {\bf (c)} and {\bf (d)} are 3, 5, 10, 15, and 20-$\sigma$
above the mean local background.  The dashed contours are 50, 100,
150, and 200-$\sigma$ above the mean local background.}
\end{figure}


\begin{figure}
\figurenum{11}
\caption{{\bf (a)} Curtis Schmidt \ha\ PDS scan of the LMC Spur
(Courtesy R. Kennicutt). {\bf (b)} \rosat\ PSPC image of the LMC Spur
in the R4 -- R7 (0.44 -- 2.04~keV) energy band.  {\bf (c)} and {\bf
(d)} The same as {\bf (a)} and {\bf (b)} with X-ray contours overlaid.
The solid contours in {\bf (c)} and {\bf (d)} are 3, 5, 10, 15, and
20-$\sigma$ above the mean local background.  The dashed contours are
50, 100, 150, and 200-$\sigma$ above the mean local background.}
\end{figure}


\begin{figure}
\figurenum{12}
\caption{{\bf (a)} \ha\ Sky Survey image of the LMC Bar (Gaustad et
al. 1999).  {\bf (b)} \rosat\ PSPC image of the LMC Bar in the R4 --
R7 (0.44 -- 2.04~keV) energy band.  {\bf (c)} and {\bf (d)} The same
as {\bf (a)} and {\bf (b)} with X-ray contours overlaid.  The solid
contours in {\bf (c)} and {\bf (d)} are 3, 5, 10, 15, and 20-$\sigma$
above the mean local background.  The dashed contours are 50, 100,
150, and 200-$\sigma$ above the mean local background.}
\end{figure}


\begin{figure}
\figurenum{13}
\caption{Plot of \rosat\ PSPC count ratios of the R6 + R7 (0.73 --
2.04~keV) energy band to the R4 + R5 (0.44 -- 1.21~keV) energy band
for the individual objects versus foreground absorption column.  The
nearly horizontal lines are lines of constant temperature for a 40\%
solar abundance Raymond \& Smith (1977) thermal plasma.  They
correspond to a thin, thermal plasma with temperatures ranging from
Log $({\rm {T \over K}}) = 6.1$ to Log $({\rm {T \over K}}) = 6.9$ with
an increment of 0.2 dex.  The PSPC ${\rm (R6 + R7)} \over {\rm (R4 +
R5)}$ count ratios for each object are plotted at the mid-point
between the lower and upper limits on the foreground absorption
column.  The vertical error bars show the errors on the count ratio.
The horizontal error bars show the range of the assumed foreground
absorption for each object.}
\end{figure}

\clearpage

\begin{deluxetable}{lccccl}
\tabletypesize{\scriptsize}
\tablecaption{PSPC Observations Included in the Mosaic. \label{tbl:pspc-obs}}
\tablewidth{0pt}
\tablehead{
\colhead{Sequence} & \colhead{Sets\tablenotemark{a}} & 
\colhead{$\alpha$} & \colhead{$\delta$} & \colhead{Exposure} 
& \colhead{Target} \\
\colhead{No.} & \colhead {} & \colhead{(J2000)} 
& \colhead{(J2000)} & \colhead{(sec)} & \colhead{Name} }
\startdata
400161 & 1 & 4 40 \phn 0.0 & -68 10     12.0 & \phn    2567.0 & RX J0440-6810               \\
500263 & 1 & 4 55     21.6 & -67 09 \phn 0.0 &        12273.7 & N9                          \\
500258 & 1 & 4 55     43.2 & -68 39 \phn 0.0 &        12379.1 & N 86                        \\
900320 & 2 & 4 56     33.6 & -66 28     48.0 &        30435.4 & N11                         \\
600098 & 1 & 5 00 \phn 4.8 & -66 25     48.0 &        10285.5 & REGION C                    \\
600577 & 1 & 5 00 \phn 4.8 & -66 25     48.0 & \phn    8792.9 & REGION C                    \\
500060 & 1 & 5 05     43.2 & -67 52     47.5 & \phn    3821.8 & SNR 0505-67.9               \\
300129 & 1 & 5 08 \phn 0.0 & -68 37     47.5 & \phn    3936.2 & NOVA LMC 1988 NO. 2         \\
500037 & 1 & 5 09 \phn 0.0 & -68 43     48.0 & \phn    6667.0 & N 103B                      \\
500063 & 1 & 5 09     31.2 & -67 31     11.5 & \phn    8496.8 & SNR 0509-67.5               \\
180033 & 1 & 5 13     52.8 & -69 51 \phn 0.0 & \phn    2472.0 & TOO RX J0513.9-6951         \\
500052 & 2 & 5 13     55.2 & -67 20     23.5 &        10767.6 & DEM105                      \\
900398 & 3 & 5 13     55.2 & -69 52     12.0 &        12005.6 & RX J0513.9-6951             \\
500061 & 1 & 5 19     33.6 & -69 02     24.0 & \phn    3706.4 & SNR 0519-69.0               \\
400263 & 2 & 5 20     28.8 & -71 57     36.0 &        21783.9 & LMCX-2                      \\
500053 & 1 & 5 20     48.0 & -65 28     12.0 & \phn    8120.0 & DEM137                      \\
110109 & 1 & 5 21     28.5 & -70 35     42.0 & \phm{--} 432.6 & XRT/PSPC SPEC/FLUX N132D    \\
500093 & 1 & 5 22 \phn 2.4 & -67 55     12.0 & \phn    8493.7 & N44                         \\
400154 & 1 & 5 22     26.4 & -67 58     12.0 & \phn    6375.3 & N44C/STAR 2                 \\
300126 & 1 & 5 23     50.4 & -70 00     36.0 & \phn    7420.4 & NOVA LMC 87                 \\
141507 & 1 & 5 25 \phn 0.0 & -69 38     24.0 & \phn    1296.7 & XRT/PSPC SPEC/FLUX N132D    \\
141508 & 1 & 5 25 \phn 0.0 & -69 38     24.0 & \phn    1106.1 & XRT/PSPC SPEC/FLUX N132D    \\
141519 & 1 & 5 25 \phn 0.0 & -69 38     24.0 & \phn    1074.3 & XRT/PSPC SPEC/FLUX N132D    \\
141542 & 1 & 5 25 \phn 0.0 & -69 38     24.0 & \phn    1627.3 & XRT/PSPC SPEC/FLUX N132D    \\
141543 & 1 & 5 25 \phn 0.0 & -69 38     24.0 & \phn    1464.0 & XRT/PSPC SPEC/FLUX N132D    \\
141800 & 1 & 5 25 \phn 0.0 & -69 38     24.0 & \phn    1010.1 & XRT/PSPC SPEC/FLUX N132D    \\
141937 & 1 & 5 25 \phn 0.0 & -69 38     24.0 & \phn    1880.4 & XRT/PSPC SPEC/FLUX N132D    \\
142011 & 1 & 5 25 \phn 0.0 & -69 38     24.0 & \phn    2526.6 & XRT/SPEC/FLUX N132D         \\
500141 & 2 & 5 25 \phn 2.4 & -69 38     24.0 &        11101.7 & N 132 D                     \\
500062 & 1 & 5 25     28.8 & -65 59     24.0 & \phn    5723.5 & SNR 0525-66.0               \\
500054 & 2 & 5 25     52.8 & -67 30 \phn 0.0 & \phn    7302.3 & DEM192                      \\
600099 & 1 & 5 26     24.0 & -66 13     48.0 & \phn    8580.8 & REGION E                    \\
600578 & 1 & 5 26     24.0 & -66 13     48.0 &        10228.3 & REGION E                    \\
500138 & 3 & 5 26     36.0 & -68 50     23.5 &        30604.4 & N144                        \\
400148 & 1 & 5 27     48.0 & -69 54 \phn 0.0 & \phn    5967.1 & RX J0527.8-6954             \\
400298 & 3 & 5 27     48.0 & -69 54 \phn 0.0 &        15875.0 & RX J0527.8-6954             \\
201249 & 1 & 5 27     50.4 & -65 56     24.0 & \phn    1363.1 & HD 36705                    \\
201258 & 1 & 5 27     57.6 & -65 57 \phn 0.0 & \phn    1799.7 & HD 36705                    \\
201592 & 1 & 5 27     57.6 & -65 57 \phn 0.0 & \phn    3384.9 & HD 36705                    \\
900542 & 1 & 5 28     40.8 & -66 48     35.5 & \phm{--} 965.2 & LMC4, POINTING 13           \\
180027 & 1 & 5 28     43.2 & -65 27 \phn 0.0 & \phn    2520.0 & TOO ORFEUS AB DOR           \\
201588 & 1 & 5 28     43.2 & -65 27 \phn 0.0 & \phn    2052.4 & HD 36705                    \\
200138 & 1 & 5 28     45.6 & -65 27 \phn 0.0 & \phn    1495.0 & HD 36705                    \\
200873 & 1 & 5 28     45.6 & -65 27 \phn 0.0 & \phn    1177.9 & HD 36705                    \\
200874 & 2 & 5 28     45.6 & -65 27 \phn 0.0 & \phn    2126.5 & HD 36705                    \\
200875 & 1 & 5 28     45.6 & -65 27 \phn 0.0 & \phn    1162.0 & HD 36705                    \\
200877 & 1 & 5 28     45.6 & -65 27 \phn 0.0 & \phn    1568.8 & HD 36705                    \\
900541 & 1 & 5 29 \phn 4.8 & -66 56     24.0 & \phm{--} 983.3 & LMC4, POINTING 12           \\
900540 & 1 & 5 29     31.2 & -66 42 \phn 0.0 & \phm{--} 973.7 & LMC4, POINTING 11           \\
201591 & 1 & 5 29     33.6 & -65 57 \phn 0.0 & \phn    1421.8 & HD 36705                    \\
900543 & 1 & 5 29     57.6 & -66 49     48.0 & \phn    1028.8 & LMC4, POINTING 14           \\
200692 & 1 & 5 30     45.6 & -65 54     36.0 &        34813.7 & LMC X-4 AND AB DOR          \\
900544 & 1 & 5 30     48.0 & -66 43     12.0 & \phn    2069.6 & LMC4, POINTING 15           \\
201589 & 1 & 5 30     55.2 & -65 54 \phn 0.0 & \phn    1199.3 & HD 36705                    \\
201248 & 1 & 5 31 \phn 2.4 & -65 54 \phn 0.0 & \phn    1193.0 & HD 36705                    \\
900553 & 1 & 5 31     12.0 & -66 51     36.0 & \phn    1182.0 & LMC4, POINTING 24           \\
110173 & 1 & 5 31     14.4 & -69 34     12.0 & \phn    1767.9 & XRT/PSPC PSF LMC X-1        \\
201254 & 1 & 5 31     24.0 & -65 52     12.0 & \phn    1352.6 & HD 36705                    \\
900539 & 1 & 5 31     36.0 & -66 59     24.0 & \phn    1342.7 & LMC4, POINTING 10           \\
201250 & 1 & 5 31     48.0 & -65 50     23.5 & \phn    3327.3 & HD 36705                    \\
900538 & 1 & 5 32 \phn 2.4 & -66 44     24.0 & \phn    1342.6 & LMC4, POINTING 09           \\
110176 & 1 & 5 32     12.0 & -70 08     24.0 & \phn    1547.8 & XRT/PSPC PSF LMC X-1        \\
140636 & 1 & 5 32     16.8 & -63 54     36.0 & \phn    1561.7 & XRT/PSPC BORE WOB LMC X-3   \\
300172 & 3 & 5 32     28.8 & -70 21     36.0 &        12737.8 & NOVA LMC 88 \#1             \\
400246 & 1 & 5 32     50.4 & -66 22     12.0 &        13986.7 & 4U 0532-664                 \\
201253 & 1 & 5 32     52.8 & -65 43     12.0 & \phn    2056.4 & HD 36705                    \\
900547 & 2 & 5 32     52.8 & -67 00     36.0 & \phn    1368.3 & LMC4, POINTING 18           \\
201256 & 1 & 5 33 \phn 0.0 & -65 42 \phn 0.0 & \phn    1315.0 & HD 36705                    \\
900546 & 2 & 5 33     16.8 & -66 45     36.0 & \phn    1034.4 & LMC4, POINTING 17           \\
900536 & 1 & 5 34     33.6 & -66 46     48.0 & \phm{--} 983.0 & LMC4, POINTING 07           \\
900552 & 1 & 5 34     57.6 & -66 55     12.0 & \phn    2037.9 & LMC4, POINTING 23           \\
160068 & 1 & 5 35 \phn 9.6 & -64 42     36.0 & \phn    1400.2 & XRT/PSPC BORE NOWOB         \\
900549 & 1 & 5 35     24.0 & -67 03 \phn 0.0 & \phn    1296.6 & LMC4, POINTING 20           \\
500100 & 2 & 5 35     28.8 & -69 16     11.5 &        25669.7 & SN1987A                     \\
500140 & 3 & 5 35     28.8 & -69 16     11.5 &        24250.7 & SN1987A                     \\
500303 & 1 & 5 35     28.8 & -69 16     11.5 & \phn    9157.4 & SN 1987 A                   \\
600100 & 2 & 5 35     38.4 & -69 16     11.5 &        18897.8 & REGION F                    \\
900548 & 1 & 5 35     50.4 & -66 48 \phn 0.0 & \phn    1045.5 & LMC4, POINTING 19           \\
300335 & 1 & 5 36     12.0 & -70 45 \phn 0.0 & \phn    9904.3 & 2 NEW SUPERSOFT SRCS        \\
900535 & 1 & 5 36     38.4 & -67 04     12.0 & \phm{--} 837.2 & LMC4, POINTING 06           \\
140007 & 1 & 5 36     40.8 & -64 05     23.5 & \phn    1015.1 & XRT/PSPC BORE NOWOB LMC X-3 \\
140637 & 1 & 5 36     40.8 & -64 05     23.5 & \phn    1768.7 & XRT/PSPC BORE WOB LMC X-3   \\
110175 & 1 & 5 36     52.8 & -69 49     12.0 & \phn    1686.2 & XRT/PSPC PSF LMC X-1        \\
110174 & 1 & 5 37 \phn 4.8 & -69 37     47.5 & \phn    2500.3 & XRT/PSPC PSF LMC X-1        \\
110168 & 1 & 5 37 \phn 7.2 & -69 01     48.0 & \phn    1794.0 & XRT/PSPC PSF LMC X-1        \\
141806 & 1 & 5 37 \phn 9.6 & -64 22     47.5 & \phn    1202.8 & XRT/PSPC LMC X-3 BORE       \\
140005 & 1 & 5 37     24.0 & -64 49     12.0 & \phn    1215.4 & XRT/PSPC BORE NOWOB LMC X-3 \\
140635 & 1 & 5 37     24.0 & -64 49     12.0 & \phn    1148.2 & XRT/PSPC BORE WOB LMC X-3   \\
900551 & 1 & 5 37     52.8 & -67 05     23.5 & \phm{--} 961.0 & LMC4, POINTING 22           \\
110167 & 1 & 5 38     30.0 & -69 31     19.0 & \phn    2200.4 & XRT/PSPC PSF LMC X-1        \\
500131 & 1 & 5 38     33.6 & -69 06     36.0 &        15449.0 & N157                        \\
900532 & 1 & 5 38     45.6 & -66 58     12.0 & \phn    1147.2 & LMC4, POINTING 03           \\
130001 & 1 & 5 38     55.2 & -64 04     48.0 & \phn    1364.7 & XRT/PSPC BORE NOWOB LMC X-3 \\
130002 & 1 & 5 38     55.2 & -64 04     48.0 & \phn    1675.6 & XRT/PSPC BORE WOB LMC X-3   \\
141805 & 1 & 5 38     57.6 & -64 04     48.0 & \phm{--} 996.6 & XRT/PSPC LMC X-3 BORE       \\
400078 & 1 & 5 38     57.6 & -64 04     48.0 & \phn    7220.3 & LMC X-3                     \\
110182 & 1 & 5 39 \phn 0.0 & -69 59     24.0 & \phn    1772.2 & XRT/PSPC PSF LMC X-1        \\
140632 & 1 & 5 39 \phn 4.8 & -63 50     24.0 & \phn    1400.4 & XRT/PSPC BORE WOB LMC X-3   \\
900533 & 1 & 5 39 \phn 9.6 & -67 06     36.0 & \phn    1534.9 & LMC4, POINTING 04           \\
140004 & 1 & 5 39     24.0 & -64 19     48.0 & \phn    1378.5 & XRT/PSPC BORE NOWOB LMC X-3 \\
140634 & 1 & 5 39     24.0 & -64 19     48.0 & \phn    1640.1 & XRT/PSPC BORE WOB LMC X-3   \\
120006 & 1 & 5 39     38.4 & -69 44     24.0 & \phn    1301.3 & XRT/PSPC LMC X-1            \\
120101 & 1 & 5 39     38.4 & -69 44     24.0 & \phn    1539.4 & XRT/PSPC LMC X-1            \\
400079 & 2 & 5 39     38.4 & -69 44     24.0 & \phn    5470.0 & LMC X-1                     \\
140631 & 1 & 5 39     48.0 & -63 20     24.0 & \phn    1482.3 & XRT/PSPC BORE WOB LMC X-3   \\
150044 & 1 & 5 40     12.0 & -69 19     48.0 & \phn    5151.6 & PSR 0540-69                 \\
400052 & 1 & 5 40     12.0 & -69 19     48.0 & \phn    6547.2 & PSR 0540-69                 \\
400133 & 1 & 5 40     12.0 & -69 19     48.0 & \phn    1722.0 & PSR 0540-69                 \\
110179 & 1 & 5 40     43.2 & -69 30     36.0 & \phn    1731.9 & XRT/PSPC PSF LMC X-1        \\
110170 & 1 & 5 41 \phn 0.0 & -69 57     36.0 & \phn    2088.2 & XRT/PSPC PSF LMC X-1        \\
140008 & 1 & 5 41     12.0 & -64 04     12.0 & \phn    1266.9 & XRT/PSPC BORE NOWOB LMC X-3 \\
140638 & 1 & 5 41     12.0 & -64 04     12.0 & \phn    1425.9 & XRT/PSPC BORE WOB LMC X-3   \\
141807 & 1 & 5 41     19.2 & -63 51 \phn 0.0 & \phn    1131.9 & XRT/PSPC LMC X-3 BORE       \\
160071 & 1 & 5 41     36.0 & -63 23     24.0 & \phn    1444.9 & XRT/PSPC BORE NOWOB         \\
110180 & 1 & 5 42 \phn 8.6 & -69 01     55.0 & \phn    1270.6 & XRT/PSPC PSF LMC X-1        \\
110178 & 1 & 5 42     16.8 & -69 39 \phn 0.0 & \phn    1937.7 & XRT/PSPC PSF LMC X-1        \\
110171 & 1 & 5 42     19.2 & -69 50     23.5 & \phn    1999.5 & XRT/PSPC PSF LMC X-1        \\
110169 & 1 & 5 44     26.4 & -70 22     12.0 & \phn    1846.1 & XRT/PSPC PSF LMC X-1        \\
141851 & 1 & 5 44     48.0 & -65 43     48.0 & \phn    4447.3 & XRT/PSPC UV LEAK, DELTA DOR \\
140009 & 1 & 5 45     36.0 & -63 54     36.0 & \phn    1472.4 & XRT/PSPC BORE NOWOB LMC X-3 \\
140639 & 1 & 5 45     36.0 & -63 54     36.0 & \phn    1584.0 & XRT/PSPC BORE WOB LMC X-3   \\
400012 & 1 & 5 46     45.6 & -71 09 \phn 0.0 &        15496.1 & CAL87                       \\
400013 & 1 & 5 46     45.6 & -71 09 \phn 0.0 &        14298.1 & CAL87                       \\
500259 & 1 & 5 47 \phn 9.6 & -69 42 \phn 0.0 & \phn    3918.0 & DEM 316                     \\
110177 & 1 & 5 48 \phn 9.6 & -69 37     47.5 & \phn    1900.3 & XRT/PSPC PSF LMC X-1        \\
201610 & 1 & 5 50 \phn 0.0 & -71 52     12.0 & \phn    7668.1 & RX J0549.9-7151             \\
100406 & 1 & 6 00 \phn 0.0 & -66 33     35.5 &        18345.2 & WFC BACKGROUND SEP S1       \\
900175 & 1 & 6 00 \phn 0.0 & -70 40     12.0 & \phn    5907.2 & LMC FRONT                   \\
900174 & 1 & 6 10 \phn 0.0 & -71 30 \phn 0.0 & \phn    5896.0 & LMC FRONT                   \\
\enddata 
\tablecomments {Units of right ascension are hours, minutes, and seconds, and
units of declination are degrees, arcminutes, and arcseconds.}
\tablenotetext{a}{Number of separate data sets processed for observation.}
\end{deluxetable}

\clearpage

\begin{deluxetable}{lcc}
\tabletypesize{\scriptsize}
\tablecaption{Broad Energy Band Definitions. \label{tbl:pspc-bands}}
\tablewidth{0pt}
\tablehead{
\colhead{Band Name} & \colhead{PI Channels}
    & \colhead{Energy\tablenotemark{a}~~(keV)}}
\startdata
R1  & $8-19$    & $0.11-0.284$ \\
R2  & $20-41$   & $0.14-0.284$ \\
R4  & $52-69$   & $0.44-1.01$  \\
R5  & $70-90$   & $0.56-1.21$  \\
R6  & $91-131$  & $0.73-1.56$  \\
R7  & $132-201$ & $1.05-2.04$  \\
\enddata
\tablenotetext{a} {10\% of peak response.}
\end{deluxetable}

\clearpage

\begin{deluxetable}{lcccc}
\tabletypesize{\scriptsize}
\tablecaption{Relevant Parameters of the X-ray Mosaics. \label{tbl:stats}}
\tablewidth{0pt}
\tablehead{
\colhead{Band} & \colhead{Average} & \colhead{Total Counts}
    & \colhead{Total Counts} & \colhead{Average} \\
\colhead{}     & \colhead{Exposure (ksec)} & \colhead{Observed}
    & \colhead{Background}   & \colhead{Intensity\tablenotemark{a}
\tablenotemark{b}}}
\startdata
R1  & 20.77  & 1223488 & 258289  & 252   \\
R2  & 21.15  & 1887260 & 338936  & 397   \\
R4  & 20.33  & 1314914 & 228274  & 290   \\
R5  & 20.25  & 1884291 & 108108  & 475   \\
R6  & 19.69  & 2478833 & 129499  & 646   \\
R7  & 16.92  & 1307574 & 109995  & 383   \\
\enddata
\tablenotetext {a}{Units of $10^{-6}$~\csam.}
\tablenotetext {b}{The area covered by the PSPC mosaics is 51.3 degree$^2$.}
\end{deluxetable}

\clearpage

\begin{deluxetable}{lccccc}
\tabletypesize{\scriptsize}
\tablecaption{Large Scale Diffuse X-ray Emission Regions 
\label{tbl:regions} }
\tablewidth{0pt}
\tablehead{
\colhead{Object} & \colhead{$\alpha$\tablenotemark{a}} & 
\colhead{$\delta$\tablenotemark{a}} & 
\colhead{Dimension\tablenotemark{a}} 
& \colhead{$N_{\rm H\,I}^{\rm Gal}$\tablenotemark{,b}}  
& \colhead{$N_{\rm H\,I}^{\rm LMC}$\tablenotemark{,c}} \\
\colhead{}     & \colhead{(J2000)} & \colhead{(J2000)}
    & \colhead{(pc)} & \colhead{($10^{21}$~cm$^{-2}$)} & \colhead{($10^{21}$~cm$^{-2}$)}    }
\startdata
LMC\,1                & 05 00 11  & $-$65 35 41 & 700                & 0.39 & 1.09 \\
LMC\,2                & 05 41 36  & $-$69 28 40 & 900                & 0.67 & 3.70 \\
LMC\,3                & 05 29 38  & $-$69 17 48 & 1000               & 0.62 & 1.54 \\
LMC\,4                & 05 30 59  & $-$66 47 53 & $1400 \times 1000$ & 0.58 & 0.68 \\
LMC\,5                & 05 22 05  & $-$66 07 15 & 800                & 0.52 & 1.19 \\
LMC\,6                & 04 57 48  & $-$68 40 32 & 600                & 0.70 & 1.25 \\
LMC\,10 (DEM~L~268)   & 05 37 01  & $-$68 25 20 & $1000 \times \phn 600$  & 0.62 & 1.94 \\
LMC Spur              & 05 43 15  & $-$70 07 46 & $\phn 900  \times \phn 500$  & 0.70 & 3.06 \\
LMC Bar               & 05 22 30  & $-$69 26 17 & $2700 \times \phn 900$  & 0.63 & 1.19 \\
\enddata
\tablecomments {Units of right ascension are hours, minutes, and seconds, and
units of declination are degrees, arcminutes, and arcseconds.}
\tablenotetext {a}{The coordinates and physical sizes have been taken from
Meaburn (1980) for LMC\,1 -- LMC\,6, Davies, Elliot, \& Meaburn (1976) for
LMC\,10, and measured from Figure~1 for the LMC Spur and the LMC Bar.}
\tablenotetext {b}{From Dickey \& Lockman (1990).}
\tablenotetext {c}{From Rohlfs et al. (1984).}
\end{deluxetable}

\clearpage

\begin{deluxetable}{lcccc}
\tabletypesize{\scriptsize}
\tablecaption{PSPC Count Rates of the Large Scale Diffuse Emission Regions 
\label{tbl:pspc_cnts2}}
\tablewidth{0pt}
\tablehead{
\colhead{Object} & \colhead{Region Size\tablenotemark{a}} & 
\colhead{R4 $+$ R5}          & \colhead{R6 $+$ R7}          & 
\colhead{ ${\rm (R6 + R7)} \over {\rm (R4 + R5)}$ } \\
\colhead{}       & \colhead{(arcmin$^{2}$)}               & 
\colhead{(counts~s$^{-1}$)} & \colhead{(counts~s$^{-1}$)} & 
\colhead{} 
}
\startdata
LMC\,1  &  \phn 954  &  $0.018 \pm 0.015$ & $0.006 \pm 0.0004$ & 
$0.324 \pm 0.281$ \\ 
LMC\,2  &  1879      &  $1.201 \pm 0.382$ & $1.239 \pm 0.359$ & 
$1.032 \pm 0.444$ \\ 
LMC\,3  & 2459       &  $0.918 \pm 0.248$ & $0.584 \pm 0.266$ &
$0.637 \pm 0.337$ \\
LMC\,4  &  5144      &  $0.499 \pm 0.964$ & $0.262 \pm 0.753$ & 
$0.525 \pm 1.817$ \\ 
LMC\,5  &  \phn 883  &  $0.072 \pm 0.183$ & $0.066 \pm 0.099$ & 
$0.914 \pm 2.697$ \\ 
LMC\,6  & \phn 825   &  $0.016 \pm 0.002$ & $0.004 \pm 0.009$ & 
$0.241 \pm 0.533$ \\ 
LMC\,10 (DEM~L~268) &  1414      &  $0.589 \pm 0.400$ & $0.301 \pm 0.358$ & 
$0.510 \pm 0.700$ \\ 
LMC Spur &  1347     &  $0.841 \pm 0.373$ & $0.622 \pm 0.309$ & 
$0.739 \pm 0.492$ \\ 
LMC Bar &  5403      &  $1.318 \pm 1.474$ & $0.846 \pm 1.022$ & 
$0.641 \pm 1.057$ \\ 
\enddata
\tablenotetext {a} {Angular size of the region from which the source
counts were extracted.}
\end{deluxetable}

\clearpage

\begin{deluxetable}{lccccccc}
\tabletypesize{\scriptsize}
\tablecaption{Physical Conditions of the Hot Gas 
\label{tbl:derived}}
\tablewidth{0pt}
\rotate
\tablehead{
\colhead{Object}                  & \colhead{$l$}                 & 
\colhead{$kT^{\rm a}$}            & \colhead{$F_X^{\rm a, b}$}    & 
\colhead{$L_X^{\rm a, b, c}$}     & \colhead{$\Lambda^{\rm b}$}   & 
\multicolumn{2}{c}{$n_e^{\rm d}$} \\
\colhead{}                        & \colhead{(pc)}                & 
\colhead{(keV)}                   & 
\colhead{($10^{-11}$~erg~cm$^{-2}$~s$^{-1}$)}                     & 
\colhead{($10^{36}$~erg~s$^{-1}$)}                                & 
\colhead{($10^{-24}$~erg~cm$^{3}$~s$^{-1}$)}                      & 
\multicolumn{2}{c}{(cm$^{-3}$)} \\ 
\cline{7-8} \\ 
\colhead{} & \colhead{} & \colhead{} & \colhead{} & 
\colhead{} & \colhead{} & \colhead{$f = 0.5$} & \colhead{$f = 1.0$} }
\startdata
%
LMC\,1   & \phn  400 & 0.19 -- 0.21 & \phn 0.07     --  \phn 0.10 & 
\phn 0.21 -- \phn  0.30 & \phn  6.00 -- \phn 7.14   & 0.005 -- 0.007 & 0.004 -- 0.005 \\
LMC\,2   & \phn  500 & 0.43 -- 0.49 &     10.0 \phn --      13.15 &      
29.91 -- 39.34          & 12.61 -- 13.57            & 0.028 -- 0.033 & 0.020 -- 0.024 \\
LMC\,3   & \phn  600 & 0.32 -- 0.36 & \phn 3.6 \phn --  \phn 4.87 &      
10.77 -- 14.58          & \phn  9.91 -- 10.86       & 0.015 -- 0.018 & 0.011 -- 0.013 \\
LMC\,4   &      1060 & 0.29 -- 0.33 & \phn 1.40     --  \phn 1.93 & 
\phn 4.19 -- \phn  5.77 & \phn  9.47 -- 10.35       & 0.005 -- 0.006 & 0.003 -- 0.004 \\
LMC\,5   & \phn  400 & 0.54 -- 0.60 & \phn 0.23     --  \phn 0.28 & 
\phn 0.69 -- \phn  0.83 & 14.49 -- 14.53            & 0.006 -- 0.007 & 0.004 -- 0.005 \\
LMC\,6   & \phn  440 & 0.15 -- 0.16 & \phn 0.13     --  \phn 0.20 & 
\phn 0.39 -- \phn  0.61 & \phn  3.82 -- \phn 3.84   & 0.009 -- 0.012 & 0.007 -- 0.009 \\
LMC\,10 (DEM~L~268)  & \phn  480 & 0.24 -- 0.26 & \phn 3.5 \phn --  \phn 4.69 & 
10.47 -- 14.03          & \phn  7.89 -- \phn 8.68   & 0.024 -- 0.030 & 0.017 -- 0.021 \\
LMC Spur & \phn  460 & 0.29 -- 0.33 & \phn 6.8 \phn --  \phn 9.77 &      
20.34 -- 29.24          & \phn  9.47 -- 10.35       & 0.032 -- 0.041 & 0.023 -- 0.029 \\
LMC Bar  & \phn  775 & 0.34 -- 0.39 & \phn 4.4 \phn --  \phn 5.45 &      
13.16 -- 16.30       &      10.35 -- 11.41          & 0.010 -- 0.011 & 0.007 -- 0.008 \\
\enddata
\tablecomments {The range in values for the derived physical conditions of the
hot gas correspond to the lower and upper limits on the foreground absorption
discussed in \S4.1.  The lower limit of the foreground absorption column, $N_{H\,I}^{Gal}
+ N_{H\,I}^{LMC}$, corresponds to the upper limits on $kT$ and $\Lambda$ and the lower
limits on $F_X$, $L_X$, and $n_e$.  The upper limit of the foreground absorption column, 
$2 N_{H\,I}^{Gal} + N_{H\,I}^{LMC}$, corresponds to the lower limits on $kT$ and $\Lambda$ 
and the upper limits on $F_X$, $L_X$, and $n_e$.}
\tablenotetext {a}{Measured for a Raymond-Smith (1977) thermal plasma 
with 40\% solar abundance.}
\tablenotetext {b}{Measured in the PSPC R4 to R7 energy band 
(0.44 -- 2.04~keV).}
\tablenotetext {c}{Assuming a distance of 50~kpc to the LMC.}
\tablenotetext {d}{Derived using an assumed path length of $l$ 
and the angular size given in Table~5.}
\end{deluxetable}

\clearpage

\begin{deluxetable}{lcc}
\tabletypesize{\scriptsize}
\tablecaption{X-ray Emission in the PSPC Mosaic 
\label{tbl:xraybudget}}
\tablewidth{0pt}
\tablehead{ \colhead{Source} & \colhead{Count Rate} & \colhead{Percentage\tablenotemark{a}} \\
\colhead{} & \colhead{(counts s$^{-1}$)} & \colhead{}  
}
\startdata
Foreground \& Background Sources                     &  &                           \\
Projected Toward the LMC                             &  &                           \\
\phm{-----} AGN \& Background Galaxies
& \phm{--\,}  0.8\tablenotemark{b}                      &                           \\
\phm{-----} Foreground Stars                        
& \phn       11.3\tablenotemark{b}                      &                           \\
                                                     &  &                           \\
Unclassified Point Sources\tablenotemark{c}                          
& \phm{--\,}  7.6\tablenotemark{b}                      &                           \\
                                                     &  &                           \\
LMC Sources                                          &  &                           \\
\phm{-----} X-ray Binaries \& Supersoft Sources
& \phn       53.6\tablenotemark{b}                      & \phn 41\%                           \\
\phm{-----} SNRs
& \phn       26.9\tablenotemark{b}                      & \phn 21\%                           \\
\phm{-----} Superbubbles\tablenotemark{d}
& \phm{--\,}  2.6\tablenotemark{e}                      & \phn \phn 2\%                       \\  
\phm{-----} Supergiant Shells\tablenotemark{f}
&  \phm{--\,}  8.2\tablenotemark{e}                     & \phn \phn 6\%                       \\
\phm{-----} Diffuse Gas in the Field                 &  &                           \\
\phm{-------} LMC Spur \& LMC Bar
&  \phm{--\,}  3.4\tablenotemark{e}                     & \phn \phn 3\%                       \\
\phm{-------} Anonymous Sources                                                             
& \phn   34.7\tablenotemark{e}                          & \phn 27\%                           \\
Total X-ray emission from the LMC                     &  129.4  & 100\%                   \\
\enddata
\tablenotetext{a}{Percentages are only calculated for sources that are known to
be LMC sources.}
\tablenotetext{b}{From Haberl \& Pietsch (1999).}
\tablenotetext{c}{These objects may be associated with the LMC.}
\tablenotetext{d}{This includes emission from 30 Doradus.}
\tablenotetext{e}{Measured in the 0.1 -- 2.4~keV energy band.}
\tablenotetext{f}{LMC\,1 -- LMC\,6, and LMC\,10.}
\end{deluxetable}

\end{document}